%% file: main.tex
\documentclass[conference]{IEEEtran}
\IEEEoverridecommandlockouts
\usepackage{xparse}
\usepackage{cite}
\usepackage{amsmath,amssymb,amsfonts}
\usepackage{algorithmic}
\usepackage{subcaption}
\usepackage{graphicx}
\usepackage{textcomp}
\usepackage{xcolor}
\usepackage{booktabs}
\usepackage{multirow}
\usepackage{cleveref}
\usepackage{tcolorbox}
\usepackage{threeparttable}
\usepackage{balance}

\input{preface.tex}

\begin{document}

\title{On the Effect of Transitivity and Granularity on Vulnerability Propagation in the Maven Ecosystem}


\author{\IEEEauthorblockN{
  Amir M. Mir\IEEEauthorrefmark{1},
  Mehdi Keshani\IEEEauthorrefmark{2},
  Sebastian Proksch\IEEEauthorrefmark{3}}
\IEEEauthorblockA{
\textit{Delft University of Technology}\\
Delft, The Netherlands \\
Email: \IEEEauthorrefmark{1}s.a.m.mir@tudelft.nl,
\IEEEauthorrefmark{2}m.keshani@tudelft.nl,
\IEEEauthorrefmark{3}s.proksch@tudelft.nl}
}

\maketitle

\thispagestyle{plain}
\pagestyle{plain}

\begin{abstract}
Reusing software libraries is a pillar of modern software engineering.
In 2022, the average Java application depends on 40 third-party libraries.
Relying on such libraries exposes a project to potential vulnerabilities and may put an application and its users at risk.
Unfortunately, research on software ecosystems has shown that the number of projects that are affected by such vulnerabilities is rising.
Previous investigations usually reason about dependencies on the dependency level, but we believe that this highly inflates the actual number of affected projects.
In this work, we study the effect of transitivity and granularity on vulnerability propagation in the Maven ecosystem.
In our research methodology, we gather a large dataset of 3M recent Maven packages.
We obtain the full transitive set of dependencies for this dataset, construct whole-program call graphs, and perform reachability analysis.
This approach allows us to identify Maven packages that are actually affected by using vulnerable dependencies. 
Our empirical results show that: (1) about 1/3 of packages in our dataset are identified as vulnerable if and only if all the transitive dependencies are considered. (2) less than 1\% of packages have a reachable call path to vulnerable code in their dependencies, which is far lower than that of a naive dependency-based analysis. (3) limiting the depth of the resolved dependency tree might be a useful technique to reduce computation time for expensive fine-grained (vulnerability) analysis.
We discuss the implications of our work and provide actionable insights for researchers and practitioners.
\end{abstract}


\section{Introduction}
Software reuse is one of the best practices of modern software development~\cite{krueger1992software}.
Developers can easily access reusable libraries through the online open-source repositories of popular package management systems such as \tool{Maven}, \tool{NPM}, or \tool{PyPi}.
\tool{Snyk} reports that a prolific number of libraries is used in projects (40 for the average Java project) and that security vulnerabilities have steadily increased over the past few years in software ecosystems such as \tool{Maven} and \tool{NPM}~\cite{snyk21}.
While reusing libraries can substantially reduce development efforts, research has shown that it may pose a security threat~\cite{pham2010detection} and that many applications rely on libraries that may contain known security vulnerabilities~\cite{cadariu2015tracking}.
Lauinger et al.~\cite{lauinger2018thou} found that 37\% of websites in top Alexa domains have at least one vulnerable JavaScript library.
Once fixed, developers need to update their dependencies to use the new version, however, researchers have found that developers often keep outdated dependencies, making their applications vulnerable to attacks and exploits~\cite{kula2018developers}.
A lack of awareness regarding available updates, added integration efforts, and possible compatibility issues might represent factors that lead to this phenomenon.

Timing is crucial.
The Heartbleed vulnerability, a security bug in the OpenSSL library that was introduced in 2012, remained unnoticed until April 2014~\cite{durumeric2014matter}.
An Apache Log4j vulnerability was discovered end of 2021 that affected around 35K Java projects, which propagated to around 8\% of the complete Maven ecosystem~\cite{log4jvul}. 
These examples show that it is crucial to release fixes timely to not give attackers a chance to develop exploits.
We also need to gain a better understanding of how fast vulnerabilities are discovered, how they affect an ecosystem, and how long it takes until a fix is available.

In recent years, a number of studies investigate the impact of security vulnerabilities and their propagation in the software ecosystems~\cite{decan2018impact,alfadel2021empirical,zimmermann2019small ,liu2022demystifying}.
The reasoning of these studies is limited to package-level analysis: they consider a project vulnerable if any (transitive) dependency contains a known vulnerability.
However, a package-level analysis cannot detect whether a client application actually uses the vulnerable piece of code, which can cause false results.
Recent works~\cite{zapata2018towards, nielsen2021modular} have overcome this limitation by performing fine-grained analysis of dependency relations in call graphs, which, as a result, increases the precision of vulnerability detection.
However, due to the computational cost of such analysis, these papers have only considered a limited number of projects.

In this paper, we want to investigate both dimensions at once to understand how vulnerabilities propagate to projects in the Maven ecosystem.
There is a trade-off to be made between the extent of the ecosystem coverage and the precision of the analysis, so we will investigate the effect of two opposing forces:
\emph{transitivity} (direct vs. transitive dependencies) will substantially increase the search space, while a lower \emph{granularity} (package-level vs. method-level) has the chance to improve precision.
We will answer the following research questions for the \tool{Maven} ecosystem:

\begin{description}
\item[RQ1] How are security vulnerabilities distributed in Maven?
\item[RQ2] How does vulnerability propagation differ for package-level and method-level analyses?
\item[RQ3] How do vulnerabilities propagate to root packages?
\item[RQ4] Is it necessary to consider all transitive dependencies?
\end{description}

By answering the formulated research questions, we aim to provide new insights on the potential impact of software vulnerabilities in the Maven ecosystem.
Different from similar studies on the NPM and PyPi ecosystems \cite{alfadel2021empirical, decan2018impact, liu2022demystifying}, our research methodology for RQ2-4 is based on both \textit{coarse} and \textit{fine-grained} analysis. Specifically, from the transitivity perspective, we will investigate how vulnerabilities propagate to Maven projects by going from direct dependencies to transitive ones.
Additionally, we will investigate the difference between coarse-grained analysis (i.e., package-level) and find-grained analysis (i.e., method level) in vulnerability propagation.
To answer the above RQs, we have gathered a large dataset of 3M Maven projects and 1.3K security reports.

\rev{
Our main empirical findings shows that, (1) transitivity has a substantial impact on the vulnerabilities propagation in Maven.
Of 1.1M vulnerable projects, only 31\% have known vulnerabilities in their \textit{direct} dependencies.
(2) The level of granularity is prominent when studying vulnerability propagation in the ecosystem.
Only 1.2\% of 1.1M transitively affected projects are \textit{actually} using vulnerable code in their dependencies.
(3) Among popular Maven projects, a vulnerability may impose higher security risk to other dependent projects if call-graph based analysis is considered.
(4) Limiting the maximum considered depth of transitive dependencies can be useful to reduce the cost of computationally-expensive, fine-grained analyses.
A slight decrease in the recall of an analysis can be traded off for a reduced computation time.
}

\smallskip
\noindent
Overall, this paper presents the following main contributions:
\begin{itemize}
\item We compile a public dataset for Maven that allows to study vulnerability propagation in Maven.
\footnote{https://doi.org/10.5281/zenodo.7540492}
\item We combine insights from previous works and closely investigate 1) a substantial part of the Maven ecosystem 2)~using method-level analysis.
\item We propose a differentiated view on transitivity that considers the distance of dependencies to an application. 
\end{itemize}


\section{Related Work}\label{sec:related-work}
\paragraph{Software Ecosystem Analysis}
Different characteristics of software ecosystems have been studied over the past decade. In 2016, Wittern et al.~\cite{wittern2016look} studied the evolution of the NPM ecosystem from two perspectives: (1) the growth and development activities (2) the popularity of packages. They found that package dependencies have increased by 57.9\% from 2011 to 2015.
Kikas et al.~\cite{kikas2017structure} proposed a network-based approach for studying the ecosystems of JavaScript, Ruby, and Rust. Their study shows that the growth of dependency networks for JavaScript and Ruby. Also, the removal of a single package can affect more than 30\% of projects in the ecosystem.
Decan et al.~\cite{decan2019empirical} conducted an empirical analysis of the similarities and differences between the evolution of seven different software ecosystems. Their finding shows that the package dependency network grows over time, both in size and number of updates. Also, a small number of packages account for most of the package updates.
Wang et al.~\cite{wang2020empirical} conducted an empirical study on the usages, updates, and risks of third-party libraries in the Java ecosystem. The study found that 60.0\% libraries have at most 2\% of their APIs called across projects.
Chowdhury et al.~\cite{chowdhury2021untriviality} conducted an empirical study to understand the triviality of trivial JavaScript packages. By considering the project and ecosystem usage, they found that removing one trivial package can affect up to 29\% of the entire NPM ecosystem.

Different from the aforementioned work, our work provides a new perspective on vulnerability propagation in Maven by considering the effect of both transitivity and granularity.

\begin{figure}
	\begin{center}
		\includegraphics[width=.8\linewidth]{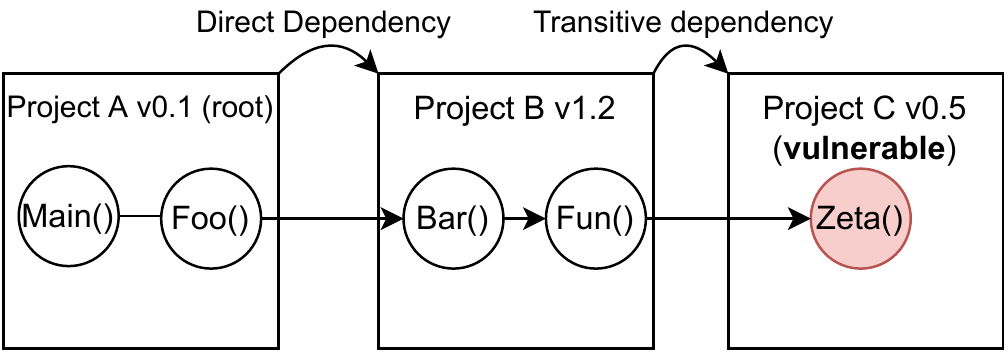}
	\end{center} \caption{A toy example that shows a root project is transitively affected by a vulnerable dependency}
	\label{fig:toy_example}
\end{figure}

\paragraph{Impact of vulnerabilities on software ecosystems}
In recent years, researchers have been studied the potential impact of security vulnerabilities in evolving software ecosystems. One of the earliest works is the master thesis of Hejderup~\cite{hejderup2015dependencies}. By considering 19 NPM packages, he studied how many dependent packages are infected by a vulnerability and how long it takes to release a fix after the publication of a security bug. 
Decan et al.~\cite{decan2018impact} studied the impact of security vulnerabilities on the NPM dependency network. Their study shows that approximately 15\% of vulnerabilities are considered high risk as they are fixed after their publication date.
Zimmermann et al.~\cite{zimmermann2019small} studied security threats in the NPM ecosystem. They found that a small number of JavaScript packages could impact a large portion of the NPM ecosystem. This implies that compromised maintainer accounts could be used to inject malicious code into the majority of the NPM packages.
Pashchenko et al.~\cite{pashchenko2020qualitative} performed a qualitative study to understand the role of security concerns on developers' decision-making for updating dependencies. The study found that developers update vulnerable dependencies if they are severe and the adoption of their fix does not require substantial efforts.
Inspired by the work of Decan et al.~\cite{decan2018impact}, Alfadel et al.~\cite{alfadel2021empirical} conducted an empirical analysis of security vulnerabilities in the PyPi ecosystem. Their findings show that PyPi vulnerabilities are discovered after 3 years and 50\% of vulnerabilities are patched after their public announcement.
Recently, Liu et al.~\cite{liu2022demystifying} studied vulnerability propagation and its evolution in the NPM ecosystem by building a complete dependency knowledge graph. Among their findings, they found that 30\% of package versions are affected by neglecting vulnerabilities in direct dependencies.

Considering the mentioned empirical studies on the impact of security vulnerabilities, their research methodology is based on dependency/package-level analysis, which highly over-estimates the number of packages using vulnerable dependencies. In contrast, we analyze projects in a lower granularity, i.e., call graph level in addition to the package level.

\begin{figure*}
    \centering
    \includegraphics[width=.9\linewidth]{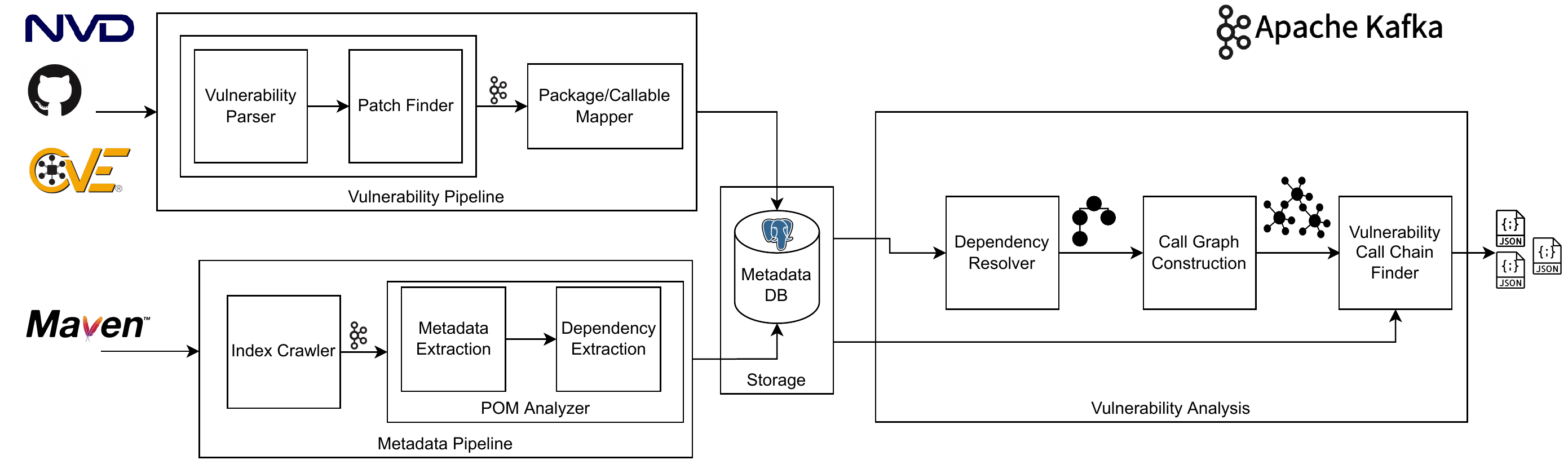}
    \caption{Overview of our data processing pipeline}
    \label{fig:dp_pipeline}
\end{figure*}

\section{Terminology}\label{sec:terms}
In this section, using Figure~\ref{fig:toy_example}, we define the terminologies that we use throughout the paper.

\begin{enumerate}
\itemsep0em
    \item A \textit{project} is a reusable software component, e.g., \code{junit}. We use Maven projects/packages interchangeably in the text.
    \item A \textit{(versioned) package} is a unique release of a project, for example, \code{junit-4.12}.
    \item A \textit{dependency} is a relation to a package whose functionalities are re-used to develop new software. Dependencies can be direct or transitive, e.g., the relation \texttt{A} $\rightarrow$ \texttt{B} is direct, while the relation \texttt{A} $\rightarrow$ \texttt{C} is transitive (through \texttt{B}).
    \item A \textit{root package} is the root of the dependency tree (e.g., an application) and (transitively) depends on other packages.
    \item A \textit{vulnerability} is a defect of software that can be exploited by attackers, e.g., to gain unauthorized system access~\cite{dowd2006art}.
    \item A \textit{call graph} is a directed control-flow graph that represents the calling relationship between methods/callables in a package. For instance, in Figure \ref{fig:toy_example}, \texttt{Bar()}, defined in Package \texttt{B}, is a callable.
    \item A \textit{vulnerable call chain} is a sequence of function calls that ends in a vulnerable callable. In Figure~\ref{fig:toy_example}, the call chain \texttt{A.Main()} $\rightarrow$ \texttt{A.Foo()} $\dots \rightarrow$ \texttt{C.Zeta()} is one such example. Also, in this example, \texttt{A.Main()} $\rightarrow$ \texttt{A.Foo()} is an \emph{internal} call as both callables are defined in Package \texttt{A}, whereas \texttt{A.Foo()} $\rightarrow$ \texttt{B.Bar()} is an \emph{external} call as \texttt{B} is a dependency of \texttt{A}.
\end{enumerate}

\begin{table}
    \centering
    \caption{List of sources for gathering vulnerability data}
    \label{tab:vuln_sources}
  \begin{tabular}{@{}lll@{}} 
				\toprule
				\textbf{Source} & \textbf{License} & \textbf{Updates} \\
    \midrule
				 National Vuln. Database (NVD)  & Public Domain  & 2 hours  \\
				GitHub Advisories  & Public Domain  & Daily  \\
				project-kb (by SAP)  & Apache License 2.0  & n/a  \\
				oss-fuzz-vulns (by Google) & CC-BY-4.0 & Daily \\
				\bottomrule
		\end{tabular}
\end{table}

\section{Approach}\label{sec:approach}
This section introduces our approach and the experimental methodology.
The overview of our data processing pipeline is shown in Figure~\ref{fig:dp_pipeline}.

\subsection{Vulnerability pipeline}
\paragraph{Vulnerability parser}
In order to create a knowledge base of vulnerability data, we gather information from various public sources (see Table~\ref{tab:vuln_sources} for details).
Each data source represents vulnerabilities in its own format and may not have complete information about a vulnerability.
Therefore, we have created a single vulnerability format that aggregates common metadata for further analysis.
The various fields of our vulnerability format are described in Table~\ref{tab:vuln_object}.
Our vulnerability knowledge base contains 1,306 security reports.

\begin{table}
    \centering
    \caption{Description of our common vulnerability format}
	\label{tab:vuln_object}
 	\begin{tabular}{@{}lp{6cm}@{}} 
				\toprule
				\textbf{JSON Field} & \textbf{Description}  \\
				\midrule
				ID  & A unique id (E.g. CVE-2018-9159)   \\
				Purls & Universal URLs that represent vulnerable packages~\cite{purl} \\
				CPE & A structured naming scheme that represents information technology systems, software, and packages~\cite{cpe}  \\
				CVSS score & A numeric value for showing the severity of software vulnerabilities from 0.0 to 10.0~\cite{cvss} \\
				CWE &  A list of software weakness types~\cite{cwe} \\
				Severity level &  Qualitative severity rating scale based on CVSS scores~\cite{cvss} \\
				Published date & The date that a vulnerability is publicly announced \\
				Last modified date & The date that a vulnerability's metadata is updated \\
				Description &  An English description of  what software systems are affected by a vulnerability \\
				References & Extra web links that provide more information about a vulnerability \\
				Patch & Links to patch information and commits \\
				Exploits & Links to how to exploit a vulnerability \\
    \bottomrule
		\end{tabular}
\end{table}
\paragraph{Patch finder}
Patch information is not always available in the references of vulnerabilities by security advisories.
Therefore, it requires manual effort to tag a reference as a patch link. Also, to find vulnerable callables/methods, we do need patch commits that show modified methods after fixing a vulnerability.

We have devised a patch finding procedure to automate the gathering of patch commits by analysing vulnerability references. 
We perform the following steps to find patch commits from references.

\begin{itemize}
	\item For GitHub, GitLab or BitBucket references, if a reference points to a commit, we directly parse the commit. In the case of pull requests, we look for the merging commit and parse it. For issues, we look for linked pull requests or commits that mention them.
	\item In references to issue trackers (Bugzilla and Jira), we look for attachments or references in the comments of an issue.
	\item If a reference points directly to a Git commit, SVN or Mercurial revisions, we parse the linked code.
\end{itemize}

After parsing a patch commit, we compute the diff of modified files in the commit. Then we create pairs of filenames and their modified line numbers. This enables us to locate modified callables in the patch commit.

\subsection{Package/callable mapper}
\paragraph{Determine vulnerable package versions}\label{subsec:det-pkg-vuln}
Considering the package-level analysis of vulnerabilities, we first identify the releases of a project that is affected by a vulnerability.
To do so, we extract and analyze vulnerability constraints in security reports.
Of all the considered vulnerability sources in Table~\ref{tab:vuln_sources}, we only use the GitHub Advisory database\footnote{https://github.com/github/advisory-database} to extract vulnerability constraints as these get reviewed by an internal security team before publication.

To explain the analysis of vulnerability constraints, consider the vulnerability constraint \texttt{$>$1.0,$<$2.0} which shows that every version between 1.0 and 2.0 is vulnerable. To compute affected releases of a project, we perform a similar approach to the previous studies~\cite{decan2018impact}, which is described as follows.
Let us denote a project and its versions/releases by $P$ and the set $V$, respectively. To find the vulnerable versions of $P$, denoted by $V_{n}$, affected by the vulnerability $V$, The package mapper (See Figure \ref{fig:dp_pipeline}) automatically performs the following steps:
\begin{enumerate}
    \item Compute the set $V$ by scraping available releases on Maven Central at the time of the request.
    \item To obtain the set $V_{n}$:
    \begin{enumerate}[noitemsep]
        \item Analyze every vulnerability constraint defined in $V$ and find affected versions if they exist in $V$,
        \item Add all affected versions to $V_{n}$, i.e., $V_{n} \subset V$.
    \end{enumerate}
\end{enumerate}
 
 To obtain dependents that are affected by the vulnerable project $P$, we simply check if dependents rely on one of the affected versions in $V_{n}$.


\paragraph{Determine vulnerable callables}
Given a vulnerability with patch information and an affected versioned packages, to identify vulnerable callables, the callable mapper automatically annotates the nodes of its call graphs with vulnerability data as follows:
\begin{enumerate}
\itemsep0em
\item Identify the last vulnerable version $P_{lv}$ and the first patched version $P_{fp}$.
\item For both $P_{lv}$ and $P_{fp}$, find files that are modified in the patch commit.
\item Locate callables whose start and end lines include the modified lines in the patched file(s) in $P_{fp}$.
\item For located callables, propagate the vulnerability to all the affected versions for which we can find the same callables. 
\end{enumerate}


\subsection{Metadata pipeline}
\paragraph{Maven index crawler}\label{subsec:dataset}
For our study, we gather versioned packages from Maven Central, which is one of the most popular and widely-used repositories of Java artifacts.
We consider packages that were released between Sep. 2021 and Sep. 2022.
The resulting dataset consists of about 3M unique versioned packages of about 200K projects.
In Maven, versioned package are differentiated by a unique Maven coordinate that consists of a ids for group, artifact, and version (i.e., \texttt{g:a:1.2.3}).

\paragraph{POM Analyzer}
Maven projects are described in a central configuration file, the \code{pom.xml}~\cite{mvnpom}.
We parse these files using the same utils that are built into Maven and extract metadata information such as release date, Maven coordinate, repository/sources URL, and the list of dependencies defined in the POM file.

\subsection{Storage}\label{subsec:storage}
The results of both vulnerability and metadata pipelines are stored in a relational SQL database. The database schema has two SQL tables for storing metadata and dependencies of versioned packages. For storing vulnerability data, there is a SQL table to store vulnerability IDs and their corresponding statement in a JSON field. Due to the space constraint, readers can refer to our replication package for more information on the database schema.

\subsection{Analyzer pipeline}
\paragraph{Dependency resolution}\label{subsec:dep-res}
To assess how a vulnerability in a Maven package affects other projects, it is necessary to reconstruct the dependency set of a versioned package.
We resolve all versioned packages that are included in our dataset using \tool{Shrinkwrap},%
\footnote{https://github.com/shrinkwrap/resolver}
a \tool{Java} library for \tool{Maven} operations.
This downloads both the \code{pom.xml} files and the .jar files of all relevant packages into the local \code{.m2} folder.
\tool{Shrinkwrap} can resolve a complete dependency set for a given coordinate.
By statically analyzing the pom files, we can reconstruct dependency trees from this dependency set, which allows us to limit the resolution and, for example, to only include dependencies up to a certain depth.

\paragraph{Call-graph construction}\label{subsec:cg-gen}
To study the effect of granularity on vulnerability propagation we perform callable-level analysis on call graphs.
We generate whole-program static call graphs for a given Maven package using \tool{OPAL}~\cite{eichberg2014software, reif2016call, opalp}, a state-of-the-art static analysis framework for Java programs.
We configure \tool{OPAL} to use a \emph{Class Hierarchy Analysis}~(CHA) for the call graph construction~\cite{grove1997call, reif2016call}, which scales well for performing a large-scale study.
We also configured \tool{OPAL} to run with an \emph{open-package assumption}~(OPA), which will treat all non-private methods as entrypoints for the analysis.
This makes conservative worst-case assumptions and produces sound call graphs~\cite{reif2016call}, which is useful for security analysis such as our vulnerable call chain analysis.

\paragraph{Identification of vulnerable call chains}
To determine whether any method of a a versioned package calls vulnerable code from one of its transitive dependencies, we need to find at least one reachable path from the method to another vulnerable method. To achieve this, we perform a Breadth-First Search (BFS) on the whole-program call graph of the versioned package plus its transitive dependencies. While traversing the graph, we compute the shortest path from the versioned package's nodes to the vulnerable node(s). Finally, we end up with a list of vulnerable call chains and their corresponding vulnerabilities.

\subsection{Implementation details \& experimental setup}
Our whole data processing pipeline (Figure \ref{fig:dp_pipeline}) is written in Java.
The pipeline has extensible components that communicate with each other either via Apache Kafka messages or through a Postgres database.
We used JGraphT for graph traversal and operations, which provides fast and memory-efficient data structures.
We ran our experiments on a Linux server (Ubuntu 18.04) with two AMD EPYC 64-Core CPUs and 512 GB of RAM. We used Docker and Kubernetes to have multiple instances of our vulnerability analyzer application to perform fine-grained analysis at a large scale. Using the above Linux machine, it took about 2 months to analyze the whole dataset with 3M versioned Maven packages.
 
\section{Empirical Results}\label{sec:empirical-results}
In this section, we present the results of our empirical study. For each RQ, we describe a motivation, the methodology used to answer the research question, and discuss the obtained results of our analysis.

\subsection{\textbf{RQ1}: How are security vulnerabilities distributed in the Maven ecosystem?}

Previous work has shown a steady increase of projects/packages in the NPM and PyPi ecosystems~\cite{decan2018impact, alfadel2021empirical}. At the same time, security vulnerabilities have become more prevalent over the past decade. As expected, an increase in the infection of projects by vulnerabilities was observed~\cite{alfadel2021empirical}. This also creates an opportunity for attackers to craft exploits. Hence, in this RQ, we are motivated to study the distribution of security vulnerabilities in our dataset from three angles: (1) the evolution of discovered vulnerabilities over time (2) how many versioned packages are affected by vulnerabilities; and (3) what are the most commonly identified types of vulnerabilities in Maven.

The results of RQ1 do not present an extensive analysis of Maven vulnerabilities.
Instead, we follow the example of previous empirical studies~\cite{decan2018impact, alfadel2021empirical} and present useful statistics from our vulnerability dataset that can inform future research.

\paragraph{Methodology}
To answer the RQ1, we follow the methodology of Alfadel et al.~\cite{alfadel2021empirical} by performing three analyses as follows. In the first analysis, we group the discovered security vulnerabilities for the Maven projects by the time they were reported. Then, we show how vulnerabilities and affected Maven projects evolve per year. Additionally, we group newly discovered vulnerabilities per severity level. This helps to quantify the threat levels in the ecosystem.

In the second analysis, given that a vulnerability can potentially affect many versioned packages, we show how vulnerable Maven versioned packages are distributed. To do so, we consider the version constraint in our dataset to identify the list of affected versions by a vulnerability.

In the third analysis, we group the most commonly identified vulnerability types in the Maven ecosystem. In our dataset, each vulnerability is associated with a \emph{Common Weakness Enumeration} (CWE), a category of software weaknesses. Finally, we count the frequency of vulnerability types to show the most common vulnerabilities in the Maven ecosystem. Similar to the first analysis, we break the analysis by severity levels to show the distribution of threat levels for each vulnerability type. 

\begin{figure}
    \centering
    \includegraphics[width=1\linewidth]{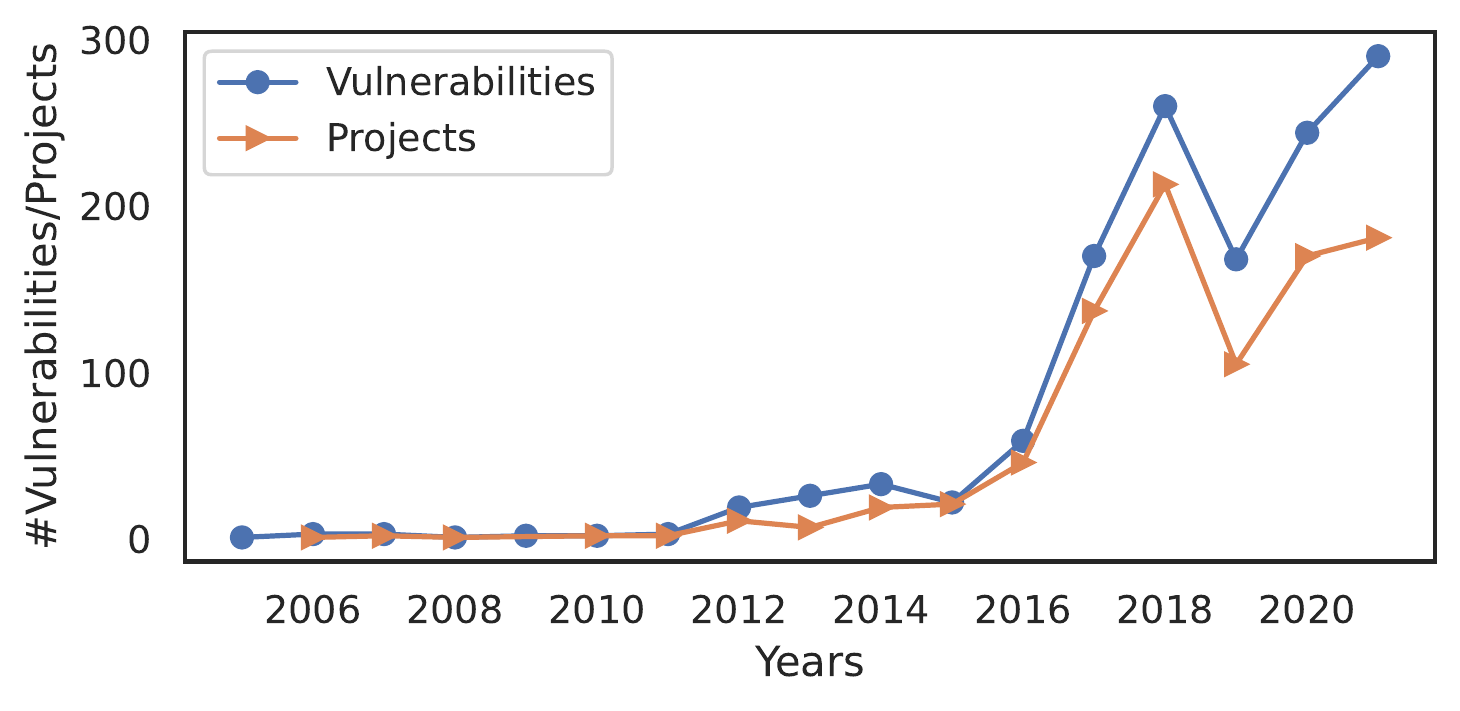}
    \vspace{-18pt}
    \caption{Vulnerability Introduction Into Maven by Year}
    \label{fig:vuln_affected_pkgs}
    \vspace{9pt}
    \includegraphics[width=1\linewidth]{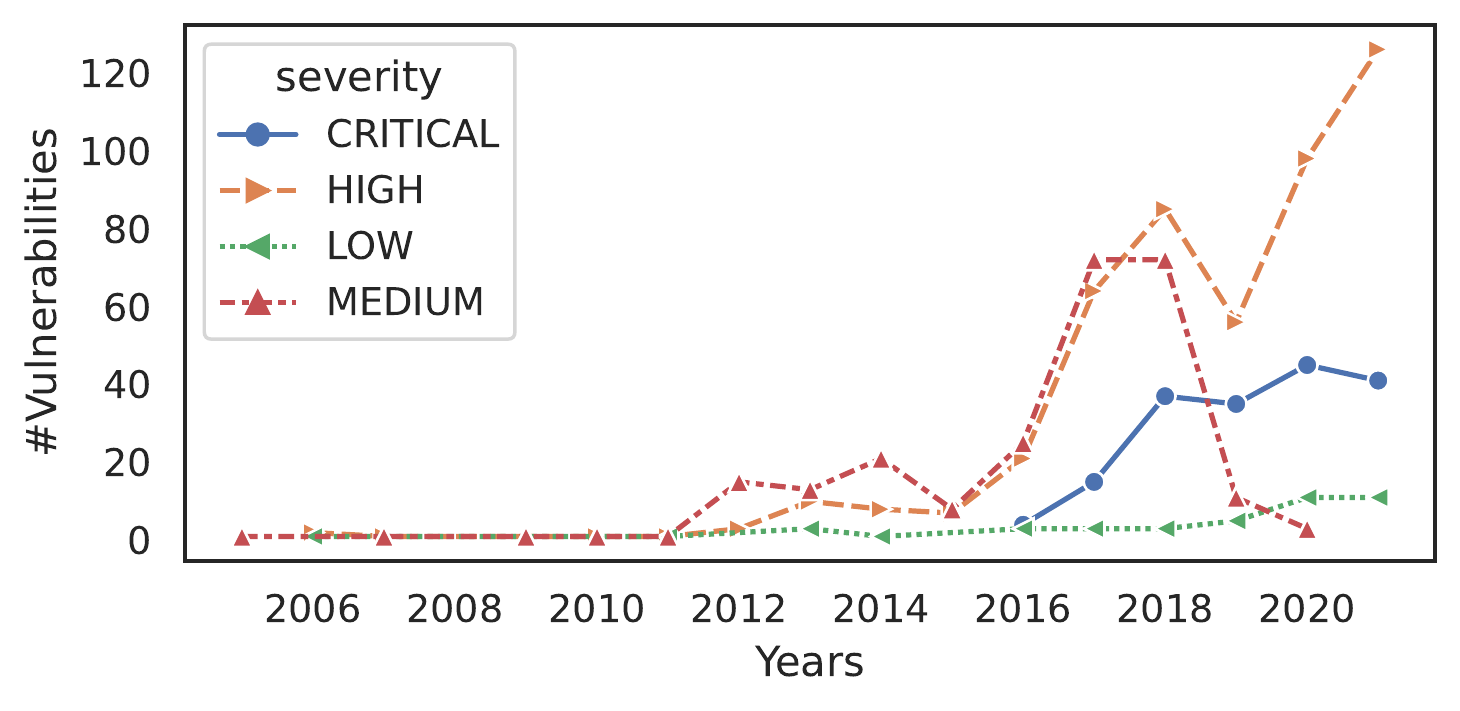}
    \vspace{-18pt}
    \caption{Vulnerability Introduction by Year and Severity}
    \label{fig:vuln_sev}
\end{figure}

\paragraph{Findings}
From Figure \ref{fig:vuln_affected_pkgs}, it can be seen that both vulnerabilities and affected projects have steadily increased in the Maven ecosystem. For instance, in 2014, 15 Maven projects were affected by vulnerabilities. In 2018, 223 Maven projects were affected, an increase of almost 15 times.

Figure \ref{fig:vuln_sev} shows the vulnerability introduction by severity level. Overall, we observe that vulnerabilities with critical and high severity levels have increased significantly over the past couple of years. Considering vulnerabilities with high severity, in 2017, 64 vulnerabilities were discovered, this number doubled in 2021, i.e., 128 vulnerabilities. This suggests that attackers may have a higher chance to craft an exploit and damage the affected software systems.

\begin{figure}
	\begin{subfigure}{0.31\columnwidth}
			\includegraphics[width=\textwidth]{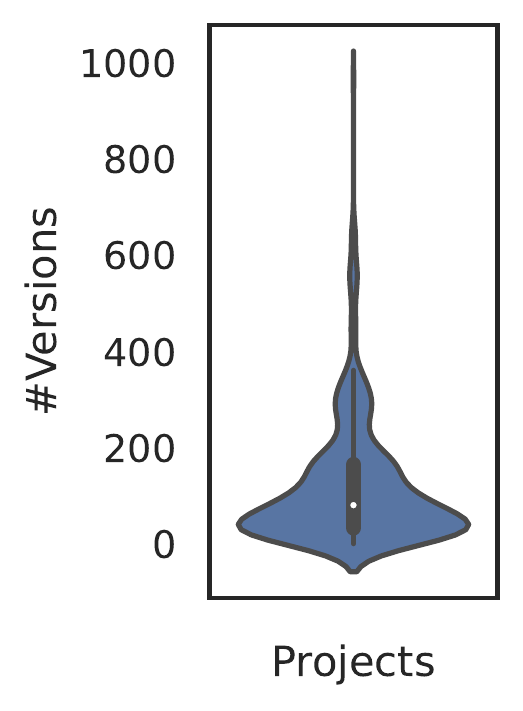}
		    \caption{Total Projects}
		    \label{fig:dist_no_vers}
	\end{subfigure}
\hfill
\begin{subfigure}{0.31\columnwidth}
	\includegraphics[width=\textwidth]{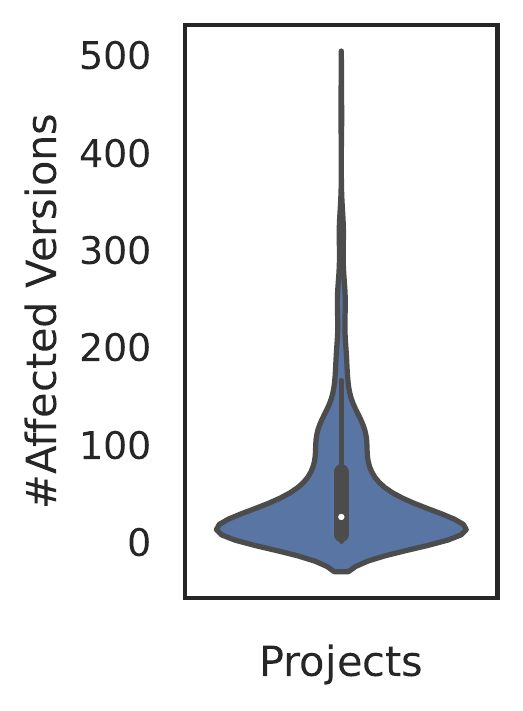}
    \caption{Vulnerable}
    \label{fig:dist_no_affected_vers}
\end{subfigure}
\hfill
\begin{subfigure}{0.31\columnwidth}
	\includegraphics[width=\textwidth]{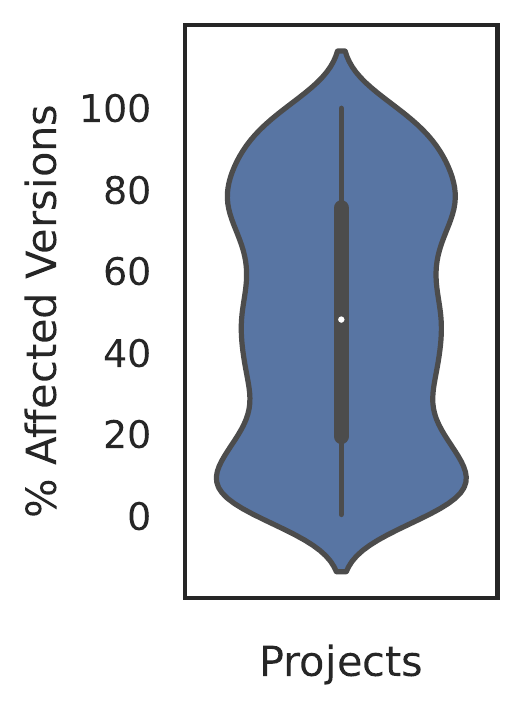}
	\caption{Vulnerable (\%)}
	\label{fig:dist_per_affec_vers}
\end{subfigure}
\caption{Vulnerability distribution among projects in the dataset}
\label{fig:dist_vuln_pkgs}
\end{figure}

From Figure \ref{fig:dist_no_vers}, it can be observed that Maven projects release often with a median of 81 unique versions.
The median Maven project also has 26 vulnerable versions, which shows that 32\% of all projects are affected considering available versions at the time of vulnerability discovery.

Our dataset contains 114 distinct software weaknesses (CWEs). Table \ref{tab:vuln_cwe_dist} shows the top 5 common software weaknesses in the Maven projects. Overall, these 5 software weaknesses account for 37\% of all the discovered vulnerabilities.
The most common vulnerability type is the \emph{deserialization of untrusted data} (CWE-502), most of which are of critical or high severity levels. This indicates a major threat to the affected Maven projects by CWE-502.


\begin{table*}
    \centering
    \caption{Top 5 most commonly found vulnerability types in Maven}
    \label{tab:vuln_cwe_dist}
    \input{tabs/top_5_maven_cves.tex}
\end{table*}

\paragraph{Comparison to the NPM and PyPi ecosystems} Similar to the existing studies on these two ecosystems~\cite{alfadel2021empirical, decan2018impact}, we also observe that security vulnerabilities have increased over time. However, Maven packages have substantially more releases, on median, 81 releases whereas PyPi, on the median, has 29 releases. Also, as expected, Maven packages have more vulnerable versions, i.e., 26, on the median, compared to 18, on the median, in PyPi.


\subsection{\textbf{RQ2}: How do vulnerabilities propagate to Maven projects considering dependency- and callable-level analyses?}

In the RQ2, we are interested in studying the effect of \textit{transitivity} and \textit{granularity} on the propagation of security vulnerabilities to Maven projects. This is different from prior similar studies~\cite{zimmermann2019small, liu2022demystifying}, which considered a project as vulnerable if one of its dependencies contain a vulnerability. This overestimates the number of affected projects and hence it may introduce false positives. Moreover, as shown in a recent study~\cite{ponta2020detection}, a project is not affected if it does not call vulnerable code in its dependencies. Specifically, from the transitivity perspective, we want to find out how many versioned packages are \textit{potentially} affected by a known vulnerability in their direct or transitive dependencies. From the granularity perspective, we want to know how many versioned packages are \textit{actually} affected by calling vulnerable code.

\paragraph{Methodology}
To answer RQ2, we perform our experiment on our Maven dataset using four distinct analysis settings:

\begin{description}
\item[$D_p(max)$:]  A package-level analysis that includes all transitive dependencies.
%
\item[$D_p(1)$:]
A package-level analysis on only direct dependencies.
%
\item[$D_m(max)$:]
A callable-level analysis that includes all  transitive dependencies.
It computes how many versioned packages are \textit{actually} affected by calling vulnerable code from their transitive dependencies.
In the whole-program call graph that we create using the OPAL framework, we mark nodes as vulnerable if modified functions in the patch commit match the signature of the node.
If there is at least a path from a node of the root project to a vulnerable node in its transitive dependencies, we consider the versioned project affected by a vulnerability.
\item[$D_m(1)$:]
A callable-level analysis that is similar to $D_m(max)$, but which only considers direct dependencies.
%
\end{description}

The subsequent sections will refer to these four defined settings.

\paragraph{Findings}
Figure~\ref{fig:vuln_affected_pkgs} shows the number of affected versioned packages considering the four described analyses in the methodology of the RQ2. Notice that the x-axis is scaled using $log_{10}$. Considering the $D_p(max)$ analysis, we observe that about $10^{6}$ versioned packages are affected by a known vulnerability in their transitive dependency set. This amounts to 40\% of versioned packages in our dataset, affected by 517 CVEs. Considering the $D_p(1)$ analysis, however, only 369K package versions are affected by using vulnerable direct dependencies, which is significantly lower than that of the $D_p(max)$ setting. This is expected as the full transitive dependency set is larger than a direct dependency set.

From Figure ~\ref{fig:vuln_pkgs_analysis_comparison}, we also observe that the callable level analysis, $D_{m}$, detects much lower vulnerable versioned packages compared to the package level analysis, $D_{p}$, i.e., $10^{4.15} \ll 10^{6}$. This is because for the $D_{m}$ setting, we perform reachability analysis to determine whether the vulnerable method in (transitive) dependencies are used whereas the $D_{p}$ setting is naive as it only checks the presence of a known vulnerability in the (transitive) dependency set. Another intriguing observation is that the set $|D_{m}(1)|=10^{3.88}$ contains more than half of the vulnerable versioned packages in the set $|D_{m}(max)|=10^{4.15}$, i.e., $|D_{m}(1) \cap D_{m}(m)|/|D_m(m)|=0.53$.

\NewDocumentCommand\distCap{}{-9pt}
\NewDocumentCommand\distPlot{}{6pt}
\begin{figure}
    \centering
    \includegraphics[width=\columnwidth]{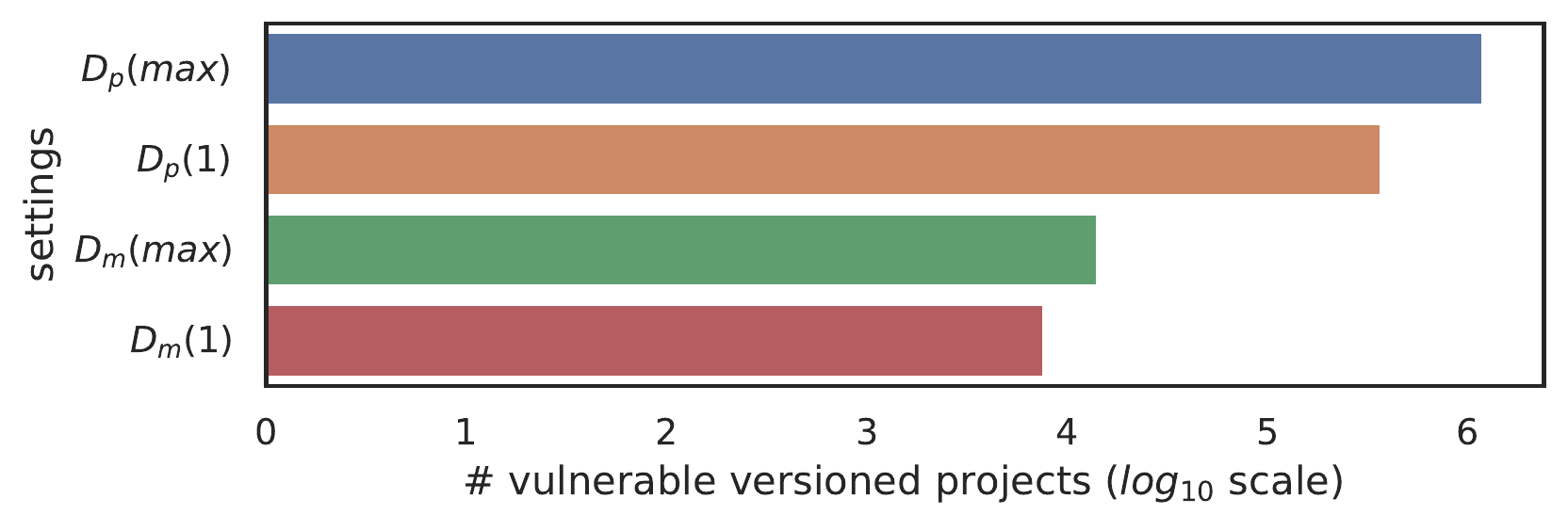}
    \vspace{\distCap}
    \caption{\#Vulnerable packages with different analysis settings}
    \label{fig:vuln_pkgs_analysis_comparison}
\vspace{\distPlot}
    \includegraphics[width=0.8\columnwidth]{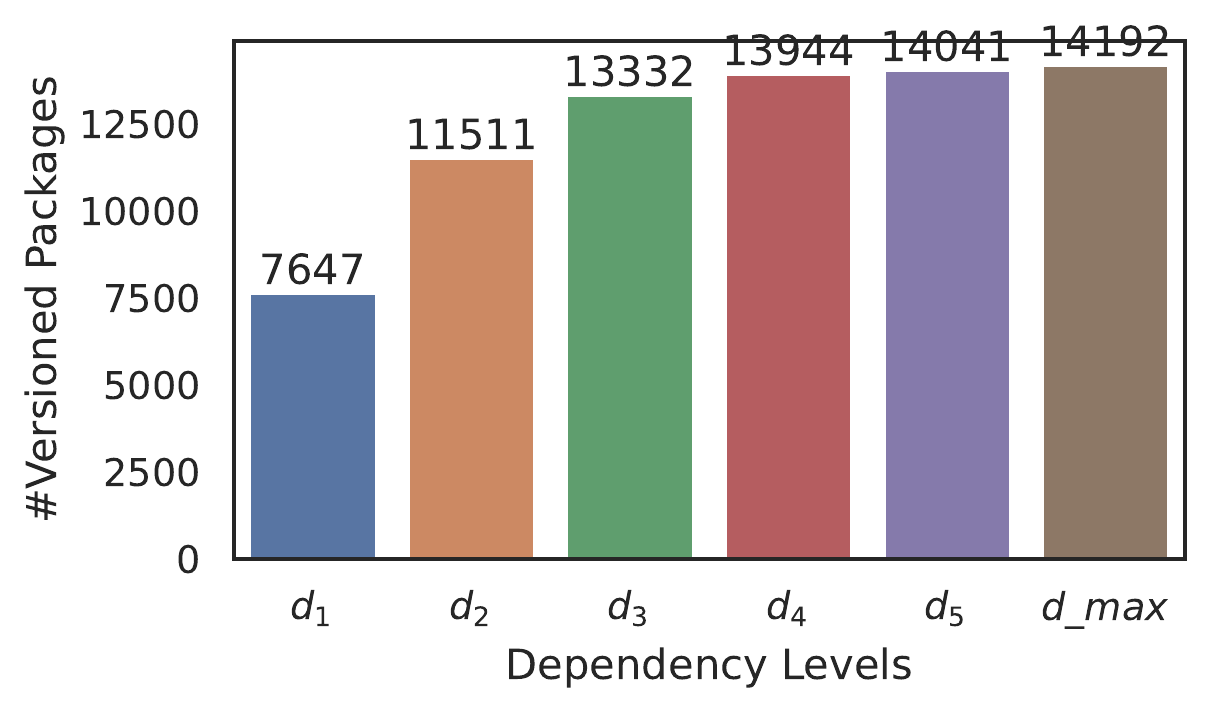}
    \vspace{\distCap}
    \caption{\#Vulnerable packages on various dependency depths}
    \label{fig:num_vuln_pkg_d_levels}
\vspace{\distPlot}
    \includegraphics[width=0.8\linewidth]{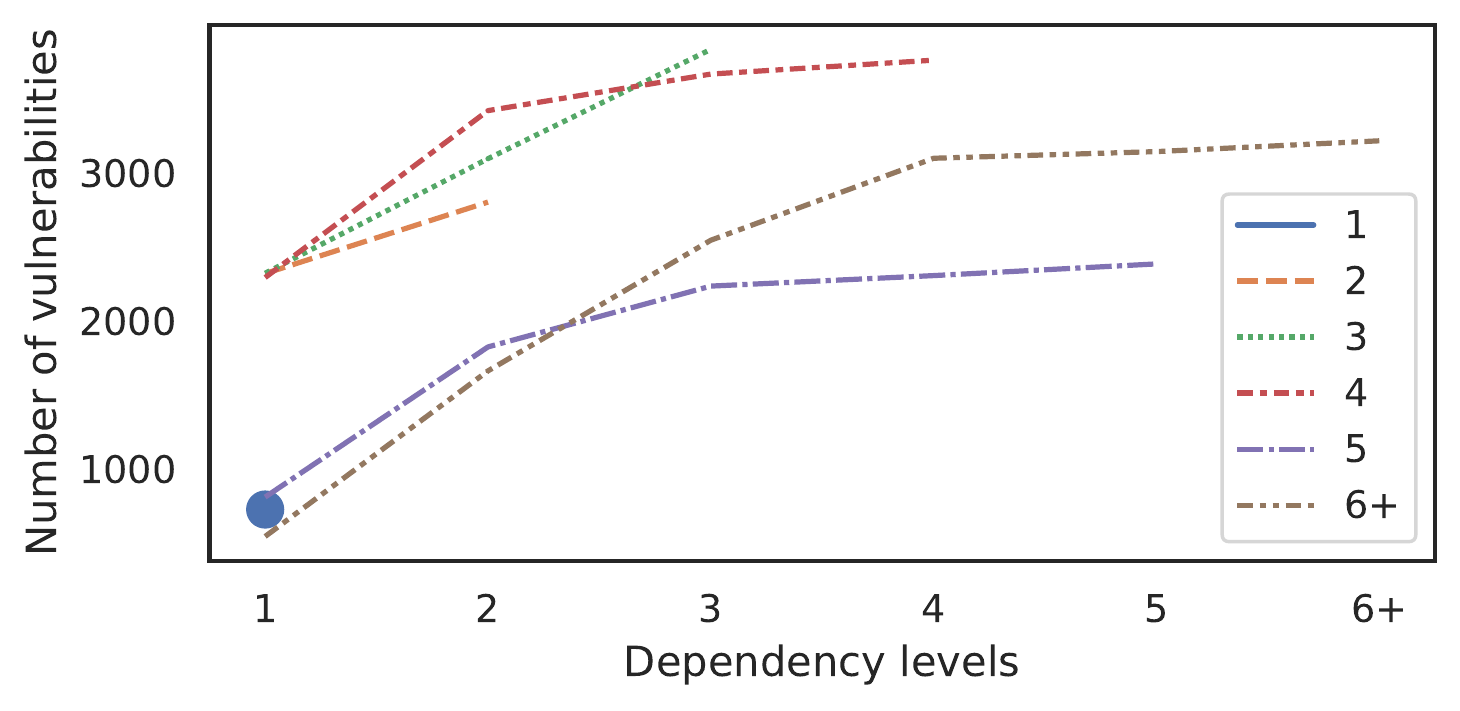}
    \vspace{\distCap}
    \caption{\#Vulnerabilities on various dependency depths}
    \label{fig:num_vuln_dep_levels}
\vspace{\distPlot}
    \includegraphics[width=0.8\linewidth]{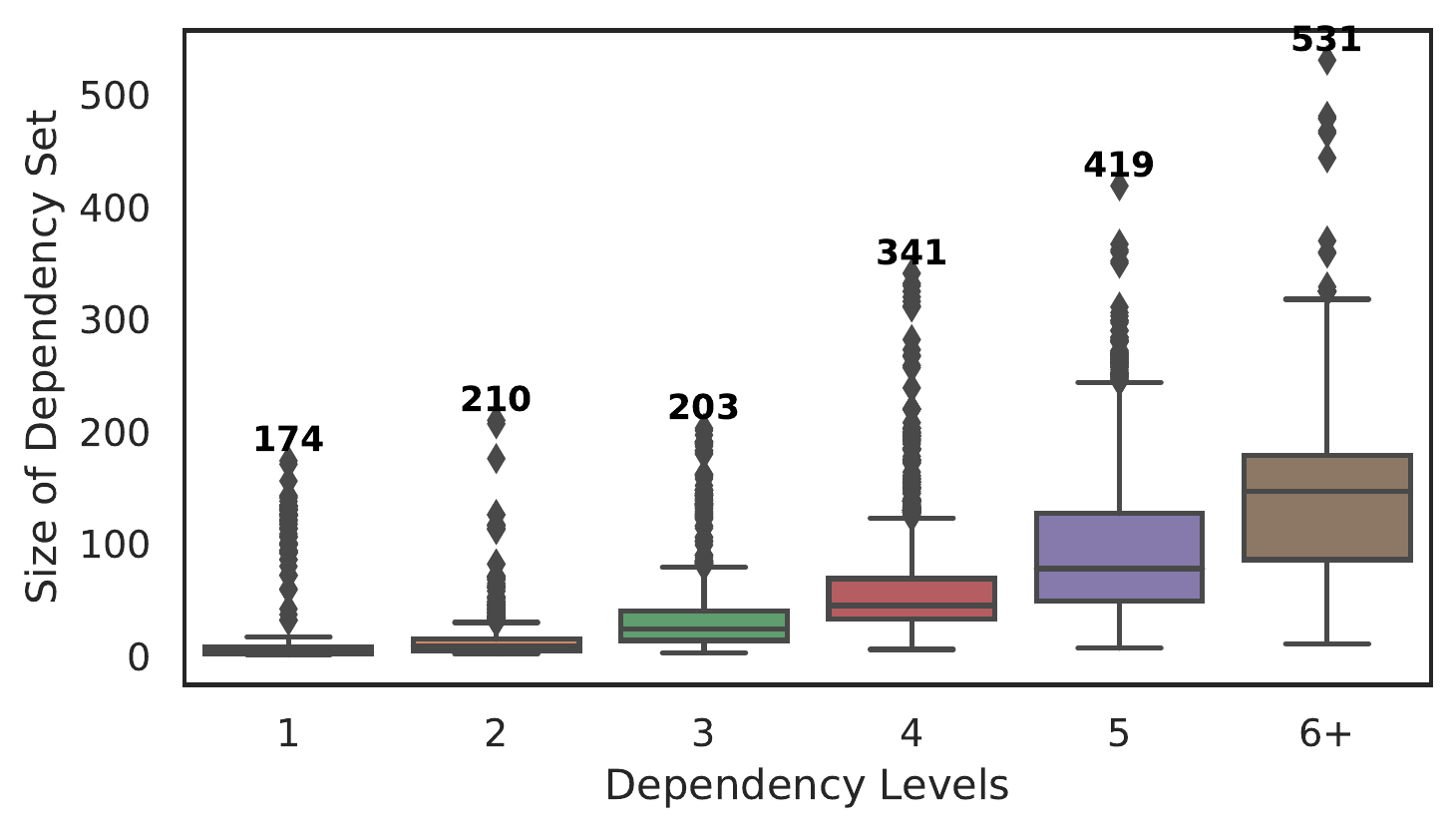}
    \vspace{\distCap}
    \caption{\#Dependencies on various dependency depths}
    \label{fig:num_deps_levels}
    \vspace{-12pt}
\end{figure}

\subsection{\textbf{RQ3}: How does the propagation of security vulnerabilities affect root packages?}

A security vulnerability in a popular package can propagate to affect many other packages in the package dependency network. This is also confirmed by a recent study~\cite{zimmermann2019small}, showing that a small number of JavaScript packages can affect a large portion of the NPM ecosystem. Therefore, we want to study how the propagation of security vulnerabilities can affect a large portion of packages and versioned packages in the Maven ecosystem. We analyze this research question from two perspectives: (1) how vulnerabilities propagate to root packages by considering transitive dependencies and (2) how vulnerabilities propagate to root packages by considering the usage of vulnerable code in dependencies.

\paragraph{Methodology}
We combine two different strategies to investigate how vulnerabilities propagate to root packages.
First, at the package level, we iterate through the full transitive dependency set of versioned packages, which is already obtained from the RQ2, i.e., the $D_{p}(max)$ setting.
We check if at least one element in the dependency set has a known vulnerability, if yes, we consider the root versioned package as vulnerable.
We list the top 10 frequent vulnerabilities that exist in the dependency trees of all the versioned projects in our dataset.
This approach overestimates the number of affected root packages, but it follows previous work~\cite{liu2022demystifying}.

Second, to analyze vulnerability propagation through vulnerable callables, we use the whole-program call graphs of versioned packages and their transitive dependencies from RQ2, i.e., $D_{m}(max)$, and then we extract known vulnerabilities, CVEs, and their corresponding vulnerable call chains.
Given these, we obtain the number of versioned packages that are actually affected by the top 10 frequent CVEs.

\begin{table*}
    \centering
    \caption{Top-10 CVEs that potentially/actually affect most package versions}
    \label{tab:top10-cves}
    \begin{threeparttable}
    \input{tabs/top_10_cves.tex}
    \begin{tablenotes}
    \item[1] The percentage of affected packages in the set $D_{p/m}(max)$. See the methodology of the RQ2 for the definition of $D_{p/m}(max)$.
  \end{tablenotes}
\end{threeparttable}
\vspace{-12pt}
\end{table*}

\paragraph{Findings}
Table \ref{tab:top10-cves} shows the top-10 CVEs that affect most versioned packages in the Maven dataset considering both dependency- and callable-level analysis.
It can be observed that the two Maven projects \texttt{jackson-databind} and \texttt{netty-codec-http} potentially affect 375,607 versioned packages in the Maven ecosystem, which is substantially higher than any other CVEs reported in Table \ref{tab:top10-cves}.
Also, even considering just the top-10 CVEs, together they already affect 786,921 versioned Maven projects, which accounts for 66.1\% of all the identified vulnerable versioned packages in the whole dataset (see $D_p(max)$ in Figure \ref{fig:vuln_pkgs_analysis_comparison}).

The results of the callable-level analysis paint a different picture.
Only 4,485 versioned Maven packages are actually affected by the top-10 CVEs.
This clearly illustrates that vulnerability analyses that only consider the package level result in a significant overestimation of vulnerable packages in the Maven ecosystem.
A second important observation is that any threat estimation will come to different conclusions, depending on whether a package-level or a callable-level granularity is being considered.
For instance, \texttt{CVE-2022-24823} (second row), accounts for 11.9\% of all potential affections, but only for 0.6\% actually affected elements.
On the other hand, \texttt{CVE-2018-1000632} (sixth row) looks much less problematic on first glance, being responsible for only 4\% of the potential affections.
However, the number of actual affections that we found is even higher than the top-1 vulnerability in the list.
This suggests that $D_p(max)$ and $D_m(max)$ do not necessarily correlate with each other when studying the vulnerability propagation and its impact on other projects.


\subsection{\textbf{RQ4}: Is considering all transitive dependencies necessary?}
The ${D_{m}(max)}$ setting can be deemed as the "best" approach to achieve high recall and precision in the vulnerability analysis. However,
to perform such analysis, one needs to compute a whole-program call graph of a versioned package plus its full transitive dependency set. This can be a very expensive task if done at the ecosystem level, i.e., a large-scale study with millions of versioned packages.
This research question investigates if it is possible to "cut-off" dependencies that are distant in the dependency tree.
Such a pruning will reduce the size of the dependency set and has a chance to speed up the fine-grained analysis at the cost of a decrease in the recall of the analysis.
We want to analyze this tradeoff.

\paragraph{Methodology}
We perform two analyses. First, we construct whole-program call graphs for all the elements of $D_{m}(max)$ and perform a reachability analysis at the dependency levels 1 to 5. This analysis produces five sets, i.e., $D_{m}(1), \dots, D_{m}(5)$. All of them are a subset of $D_{m}(max)$ (e.g., $D_{m}(2) \subset D_{m}(max)$).
In the second analysis, we find the maximum dependency depth for each versioned package in $D_m(max)$.
With this information, we iterate over the elements of $D_{m}(max)$ and count the number of reachable vulnerabilities at each dependency level until the maximum level is reached. We repeat this process for the other sets.

\paragraph{Findings}



Figure~\ref{fig:num_vuln_pkg_d_levels} shows the number of vulnerable versioned packages while performing callable-level analysis and considering different dependency levels. Using only direct dependencies, i.e., $D_{m}(1)$, 55.8\% of vulnerable versioned packages are detected comparing to $D_m(max)$. This observation is in line with the findings of RQ2 (see Figure~\ref{fig:vuln_affected_pkgs}).
Every additional layer can identify more vulnerable packages, but dependency level 3 already reaches 94\% coverage.
Cutting of at this level will result in an analysis that will miss some vulnerabilities.
While this might not be acceptable for security sensitive analyses, other analyses could leverage this finding to potentially save substantial computation time.

Figure~\ref{fig:num_vuln_dep_levels} investigate these results with a different visualization.
The different plot lines represent packages with vulnerabilities on the exact dependency level $1,2,..., 6+$.
The y-axis shows how many of the existing vulnerabilities can be found when the dependency tree is cut of at depth $d$.
As expected, vulnerable versioned packages with transitive dependencies tend to be affected by more vulnerabilities than versioned packages with only direct dependencies.
However, we see a common pattern across the different plots: the increase slows and starts to converge at dependency level 3-4.
Programs that have such deep dependency levels also have large dependency set, so for these projects, the potential saving in the analysis effort seem to be particularly beneficial.

To estimate how much computation time can potentially be reduced, we approximate the required computation time with the size of the transitive dependency set.
This is likely a lower bound, as the number of call-graph edges grows much faster than linearly.
Figure~\ref{fig:num_deps_levels} shows the distribution over the dependency set sizes for all packages in $D_m(max)$, which have a dependency tree with the exact height.
For example, the first box plot in the diagram contains all versioned packages that only have direct dependencies.
The average size of their dependency set is close to 0, whereas packages with 3 dependency levels have a median of 24 dependencies, and 6+ dependency levels even go up to a median of 147 dependencies.
Even if we only assume a linear growth in computation time, filtering the large applications to dependency level 3 would lead to an enormous analysis speed-up of about 6 times.
These large applications are usually also the limiting factor when it comes to computation timeouts or memory consumption of analyses.


\section{Discussion}\label{sec:discussion}
In this section, we discuss actionable results and provide insights from our study.

\paragraph{Granularity Matters}
When studying security vulnerabilities, granularity matters.
As shown in RQ2 and RQ3, dependency-level analysis highly overestimates the number of vulnerable packages in the Maven ecosystem.
A project is not affected if the vulnerable code/callable is never reached. This is also acknowledged in the previous related studies~\cite{decan2018impact, liu2022demystifying}. Also, for the NPM ecosystem, a similar observation was found by saying that dependency-level analysis produces many false positives~\cite{zapata2018towards}. To address this,
the callable-level analysis should be considered as it gives a more precise answer to whether a user's project actually uses the vulnerable code in its dependencies.
The results of our dependency-level analysis look worrying: we found about 175K vulnerable versioned packages in 2021 alone.
The good news is that very few seem to use vulnerable code, so most cases are \emph{actually not affected}.
The looming threat of importing vulnerabilities from open-source ecosystems is in fact much lower than popular believe.
More research is required to study this discrepancy.

\paragraph{Towards Intelligent Software Composition Analysis}
A number of free and commercial software composition analysis (SCA) tools exist that analyze the open-source components of a project for security risks and license compliance. Each of them differs widely in terms of accuracy, the quality of the vulnerability database, and the level of granularity~\cite{imtiaz2021comparative}. For instance, OWASP DC~\cite{owaspdepchk} analyzes dependency files of a project and notifies developers if known vulnerabilities are present in the project's transitive dependencies. However, as mentioned earlier, this level of granularity suffers from imprecision, and it is also not helpful for developers to better assess and mitigate the potential risk of using vulnerable dependencies in their projects. Also, free tools like GitHub's Dependabot performs package-level analysis, though its fine-grained analysis feature is in the beta state for the Python ecosystem as of this writing~\cite{ghd}. Overall, we believe that the next generation of SCA tools should have at least these core features when analyzing vulnerabilities in projects: (1) dependency depth (2) callable-level analysis (3) providing users with a detailed description of what part of their code is affected by vulnerabilities by showing, for example, vulnerable call paths and required actions to mitigate the security risk.

\paragraph{Transitivity Matters}
Transitivity matters when analyzing projects' dependencies for the presence of vulnerabilities.
Considering the results of RQ2 and RQ4, many versioned packages are affected by known vulnerabilities in the \textit{transitive} dependencies no matter the granularity level, i.e., dependency- or callable-level. For developers, this means that updating direct dependencies may not eliminate the potential security threat by a vulnerability. It is suggested for developers to use an SCA tool and integrate it into their workflow or continuous integration pipeline, which helps to frequently monitor the transitive dependencies of their projects for the presence of vulnerabilities and update them if needed. For the developers of SCA tools, it is essential to analyze the whole transitive dependency set of projects to improve the reliability of their tools. We believe that SCA tools are not practical or useful if they naively only consider direct dependencies.

\paragraph{Popularity}
Popular vulnerable projects do not necessarily have the largest impact on the ecosystem.
RQ3 shows that a security vulnerability in a popular package can potentially affect many other dependent packages.
This confirms previous results in the NPM ecosystem~\cite{zimmermann2019small}, which stated that several popular JavaScript packages (in)directly affect thousands of other packages.
However, this observation is based on a basic package-level analysis of transitive dependencies, which is not precise enough to show the \emph{true} impact of vulnerabilities in the ecosystem.
The results change, when analyzed on the method-level.
For instance, we found that a vulnerability, \texttt{CVE-2021-37137} in the popular Maven project \texttt{netty-codec-http} potentially affects 142K other packages when analyzed on the package level.
However, through a method-level analysis we only found 90 versioned packages that were actually affected.
On the other hand, the {\tt CVE-2018-1000632} in the less popular Maven project {\tt dom4j} only affects 47K other packages on the package level, but we found 1,400+ actually affected packages through a method-level analysis.
These results imply that popularity might not as good an indicator for ecosystem impact as originally thought.
Better strategies to identify the critical packages are required to protect ecosystems as a whole.

\paragraph{Expensive Analyses}
Running ecosystem-wide, fine-grained analyses is expensive.
While fine-grained analysis provides a new perspective in studying a software ecosystem, it can be very computationally expensive to analyze millions of projects.
In this study, we managed to analyze 3 million versioned Maven packages and study the effect of transitivity and granularity on vulnerability propagation in Maven. From our experience, ecosystem-wide fine-grained analysis requires costly, powerful machines and sufficient time to perform. Given the result of the RQ4, one insight that might be useful for future work is to consider a lower dependency level (e.g., 3 or 4) in call graph-based analysis assuming that a slight loss of recall/precision is acceptable. This also may potentially reduce the search space and computation time.

\section{Threats to Validity}\label{sec:threats}
In this section, we describe possible threats to the validity of the obtained results and findings and how we addressed them.

\paragraph{Dataset}
In this study, we gathered a Maven dataset that consists of 3M versioned packages over a period of one year (from 2021-2022). We chose to gather data for one year mainly for two reasons: (1) In our approach, we generate call graphs for fine-grained analysis, which can be expensive. For us, it is not computationally feasible to perform this step for the whole history of the Maven ecosystem, which has over 9.8M versioned packages~\cite{mvn22} as of this writing. (2) The main goal of this study is to show the effect of transitivity and granularity on vulnerability propagation via fine-grained analysis in Maven. Therefore, following the guidelines for empirical software engineering research~\cite{Felderer2020ContemporaryEM}, we believe that our sample size, 3M versioned packages, is sufficient to achieve the said goal.
With such a large sample size, we are very confident that our findings would also hold for the whole history of the Maven ecosystem.


\paragraph{Vulnerability mapping to package versions}
As described before, we analyze vulnerability constraints in security reports to find the affected versions of a package by a vulnerability. Based on our observation, vulnerability constraints often only specify an upper bound on the range of affected versions.
This may falsely render older releases as vulnerable. No trivial solution can address this limitation.
However, with callable-level analysis, we can check whether the vulnerable method even exists in the previous releases, which can automatically eliminate many incorrect cases.

\paragraph{Call graph analysis}
We configure \tool{OPAL} to use an Open-Package Assumption to identify entrypoints when generating call graphs.
OPA prioritizes soundness over precision, meaning that call graphs might have spurious edges, which may lead to false positives when finding vulnerable call chains. However, we argue that, for security-focused analysis, false negatives can be more expensive and dangerous. If a method is falsely identified as safe to use, it can potentially harm its users and their organizations~\cite{hagberg2021using}. In contrast, false positives prevent users to use a method and they can also be reviewed manually by security experts if the needed functionality is costly to implement. Moreover, as pointed out by Nielsen et al.~\cite{nielsen2021modular}, for security-focused applications, a few false negatives are likely more desirable than a large number of false positives. 

In addition, our call graph analysis does not consider control flow when assessing the reachability of vulnerable code or methods. This means that a false positive alarm is produced if required input to trigger the vulnerability is not provided~\cite{kang2022test}. 

\section{Summary}\label{sec:conclusion}
In this paper, we have studied the effect of transitivity and granularity on how vulnerabilities propagate to projects via fine-grained analysis in the Maven ecosystem. The methodology of our study is based on resolving transitive dependencies, building whole-program call graphs, and performing reachability analysis, which allows us to study vulnerability propagation at both dependency and callable levels. Among our findings, we found that, for security-focused applications, it is important to consider transitive dependencies regardless of the granularity level to minimize the risk of security threats. Also, with the callable-level analysis, it is possible to provide a lower bound for the analysis of vulnerability propagation in the ecosystem and also overcome the over-approximation issue of the dependency-level analysis. Overall, the implication of our results suggests that call graph-based analysis seems to be a promising direction for future studies on software ecosystems.

\section*{Acknowledgment}
The FASTEN project has received funding from the European Union's Horizon 2020 research and innovation programme under grant agreement number 825328.

\balance
\bibliographystyle{IEEEtran}
\bibliography{main}

\end{document}

%% file: preface.tex
\NewDocumentCommand\rev{+m}{#1}

\NewDocumentCommand\code{m}{{\tt #1}}
\NewDocumentCommand\tool{m}{{\small\scshape #1}}

\usepackage[noindentafter]{titlesec}
\titlespacing{\paragraph}{0pt}{\medskipamount}{4pt}
\titleformat{\paragraph}[runin]{\scshape}{\noindent}{0pt}{}[.] 

\usepackage{enumitem}
\usepackage{noindentafter}
\setlist{
	leftmargin=12pt,
	itemsep=2pt plus 1pt,
	topsep=2pt plus 1pt,
	parsep=2pt plus 1pt,
	listparindent=0pt,
}
\setlist[description]{
	font=\normalfont\itshape
}
\NoIndentAfterEnv{itemize}
\NoIndentAfterEnv{description}
\NoIndentAfterEnv{enumerate}


%% file: tabs/top_5_maven_cves.tex
\begin{tabular}{@{}lrrrrrr@{}}
			\toprule
                && \multicolumn{5}{c}{{\bf Frequency by Severity}} \\
                \cmidrule{3-7}
			{\bf Vulnerability type (CWE)} & {\bf Freq.} &  Critical & High & Moderate & Medium & Low \\
			\midrule
			Deserialization of Untrusted Data (CWE-502) & 166 & 52 &85 &17& 12 & 0 \\
			Cross-site Scripting (CWE-79) & 108 & 0 & 2 &72 & 27 & 7 \\
			Improper Input Validation (CWE-20) & 88 & 6 & 47 & 15 & 20 & 0 \\
			Improper Restriction of XML External Entity Reference (CWE-611)& 78 & 21 &32 & 10 & 11 & 4 \\
			Path Traversal (CWE-22) & 65 & 4 & 24 &18 & 19 & 0 \\
			\midrule
			Total & 505 & 83 & 190 & 132 & 89 & 11 \\
			\bottomrule
		\end{tabular}

%% file: tabs/top_10_cves.tex
\begin{tabular}{@{}llrrcrr@{}}
\toprule
   & & \multicolumn{2}{c}{{\bf Number of Packages}} && \multicolumn{2}{c}{\bf \%Proportion\tnote{1}} \\
    \cmidrule{3-4}
    \cmidrule{6-7}
    {\bf CVE ID} &                              {\bf Project} & Potentially Affected & Actually affected && $D\_p(max)$ & $D\_m(max)$ \\
\midrule
  CVE-2020-36518 & com.fasterxml.jackson.core:jackson-databind &                                  233,430 &                               1,153 &&                     19.6 &                      8.1 \\
  CVE-2022-24823 &                   io.netty:netty-codec-http &                                  142,177 &                                  90 &&                     11.9 &                      0.6 \\
  CVE-2022-24329 &          org.jetbrains.kotlin:kotlin-stdlib &                                   82,060 &                                  32 &&                      6.9 &                      0.2 \\
  CVE-2021-37137 &                        io.netty:netty-codec &                                   57,535 &                                 525 &&                      4.8 &                      3.7 \\
  CVE-2021-22569 &         com.google.protobuf:protobuf-kotlin &                                   57,095 &                                 390 &&                      4.8 &                      2.7 \\
CVE-2018-1000632 &                                 dom4j:dom4j &                                   47,820 &                               1,438 &&                      4.0 &                     10.1 \\
  CVE-2022-25647 &                   com.google.code.gson:gson &                                   47,372 &                                 171 &&                      4.0 &                      1.2 \\
   CVE-2020-8908 &                      com.google.guava:guava &                                   42,084 &                                  84 &&                      3.5 &                      0.6 \\
  CVE-2022-22965 &          org.springframework:spring-webflux &                                   38,882 &                                 572 &&                      3.3 &                      4.0 \\
  CVE-2018-20200 &                  com.squareup.okhttp:okhttp &                                   38,466 &                                  30 &&                      3.2 &                      0.2 \\
\bottomrule
\end{tabular}